\documentclass[12pt,noshowpacs,nofootinbib,notitlepage,amsmath]{revtex4-1}
\usepackage{setspace}

\allowdisplaybreaks
\usepackage{graphicx,color,amssymb}
\usepackage[colorlinks=true,citecolor=blue,linkcolor=blue,urlcolor=blue]{hyperref}
\usepackage[charter]{mathdesign}
\DeclareSymbolFontAlphabet{\mathcal}{symbols}
\DeclareSymbolFont{symbols}{OMS}{xmdcmsy}{m}{n}
\DeclareSymbolFont{largesymbols}{OMX}{xmdcmex}{m}{n}
\SetSymbolFont{symbols}{bold}{OMS}{xmdcmsy}{b}{n}
\begin{document}
\title{\color{blue} Heavy Higgs  decay to $t\bar{t} Z$ and constraints on a $750$ GeV pseudoscalar}
\author{Bob Holdom}
\email{bob.holdom@utoronto.ca}
\author{Melissa Ratzlaff}
\email{melissa.ratzlaff@gmail.com}
\affiliation{Department of Physics, University of Toronto, Toronto, Ontario M5S1A7, Canada}

\begin{abstract}
In models with multiple nondegenerate Higgs bosons, the decay chain $H/A \to A/H Z\to t\bar{t} Z$ may have a partial width comparable to the $t\bar{t}$ decay mode. We recast the ATLAS standard model $t\bar{t} Z$ measurement to put limits on the rate for this process. Limits are also set on the two Higgs doublet model at low $\tan \beta$ that are sensitive to a heavy Higgs mass as high as $\sim 750$ GeV. We then discuss the $750$ GeV diphoton excess  in terms of a pseudoscalar that also has the decays $A\to HZ$ and $A\to H^\pm W^\mp$. These decays strongly constrain the partial widths for $A\to\gamma\gamma$ and $A\to gg$ when combined with the  $t\bar t$ resonance search limits. In a benchmark model the mass of $H$ should be close to 650 GeV.
\end{abstract}
\maketitle  

\section{Heavy Higgs to $t\bar{t}Z$}
Some of the most popular extensions of the standard model (SM) include additional Higgs bosons.  When sufficiently heavy a neutral Higgs boson $(H)$ and a  pseudoscalar Higgs boson $(A)$ may have a dominant $H/A\to t\bar{t}$ final state. The masses of the  $A$ and $H$ need not be degenerate, and given a sufficient mass difference, the $H/A\to A/H Z$ decay may compete with the $H/A\to t\bar{t}$ decay. When the  lighter Higgs boson ($A$ or $H$) is above the $t\bar t$ threshold then it can be expected to have a dominant decay to $t\bar{t}$, leading to the final state $t\bar{t}Z$.  This final state has a much smaller background  than $t\bar{t}$. Also the  $HAZ$ coupling depends  only on the gauge coupling times a factor that is unity in the alignment limit of the two Higgs doublet model (2HDM). Here we determine the first limits on the Higgs cascade decay to the $t\bar{t}Z$ final state from the $8$ TeV data. In the case that the lighter of $A$ or $H$ is below the $t\bar t$ threshold, the limits from the resulting cascade decay to $b\bar bZ$ have been considered in \cite{Coleppa:2014hxa,Dorsch:2014qja,Holdom:2014boa,Dorsch:2016tab}.
   
   First we calculate the model independent limits on the $H/A\to t\bar{t}Z$ process from the ATLAS $t\bar{t}Z$ cross section measurement \cite{Aad:2015eua}. Second we use this process to constrain the masses in the 2HDM in the alignment limit. Finally we consider the implications from the $A\to t\bar{t}Z$ and $A\to H^\pm W^\mp$  decay modes when the pseudoscalar is consistent with the  $750$ GeV diphoton excess reported by ATLAS \cite{ATLASgaga,ATLASgaga2} and CMS \cite{CMS:2015dxe,CMS:2016owr}.
   
 \subsection{Model independent limits}
Both ATLAS \cite{Aad:2015eua} and CMS \cite{Khachatryan:2014ewa,Khachatryan:2015sha} have  multileptons searches which measure the SM $t\bar{t}Z$
production cross section  at $8$ TeV and also at $13$ TeV \cite{ATLAS13ttV,CMS:2016ium}.  Neither experiment sees a significant deviation from their SM expectations. The total theoretical SM $t\bar{t}Z$ cross section is $\sim 200$ fb at $\sqrt{s}=8$ TeV. Due to the small size of this cross section, the $8$ TeV data should provide useful limits on the process 
$H/A\to A/H Z\to t\bar{t}Z$. The strongest limit is given by a particular region of the ATLAS \cite{Aad:2015eua} search that has the following selection criteria:
\begin{itemize}
\item  four anti-$k_T$ jets with  $p_T>25$ GeV, 
\item three leptons with $p_T>15$ GeV,
\item on $Z$ selection within $10$ GeV of the $Z$ mass,
\item one $b$ tagged jet. 
\end{itemize}

We use CheckMATE \cite{CHECKM}  to recast this signal 
region. We also include the jet-lepton overlap removal and an approximation to the lepton isolation criteria.  We then validated our CheckMATE analysis against a sample of SM $t\bar{t}Z$ events generated with MADGRAPH \cite{Alwall:2014hca} and showered with PYTHIA 6 \cite{Sjostrand:2006za}. We use a modified 2HDM FEYNRULES model \cite{Alloul:2013bka} with HERWIG++ \cite{Bahr:2008pv} to generate our signal events. 

\begin{figure}[htb]	
\centering
\includegraphics[scale=0.41]{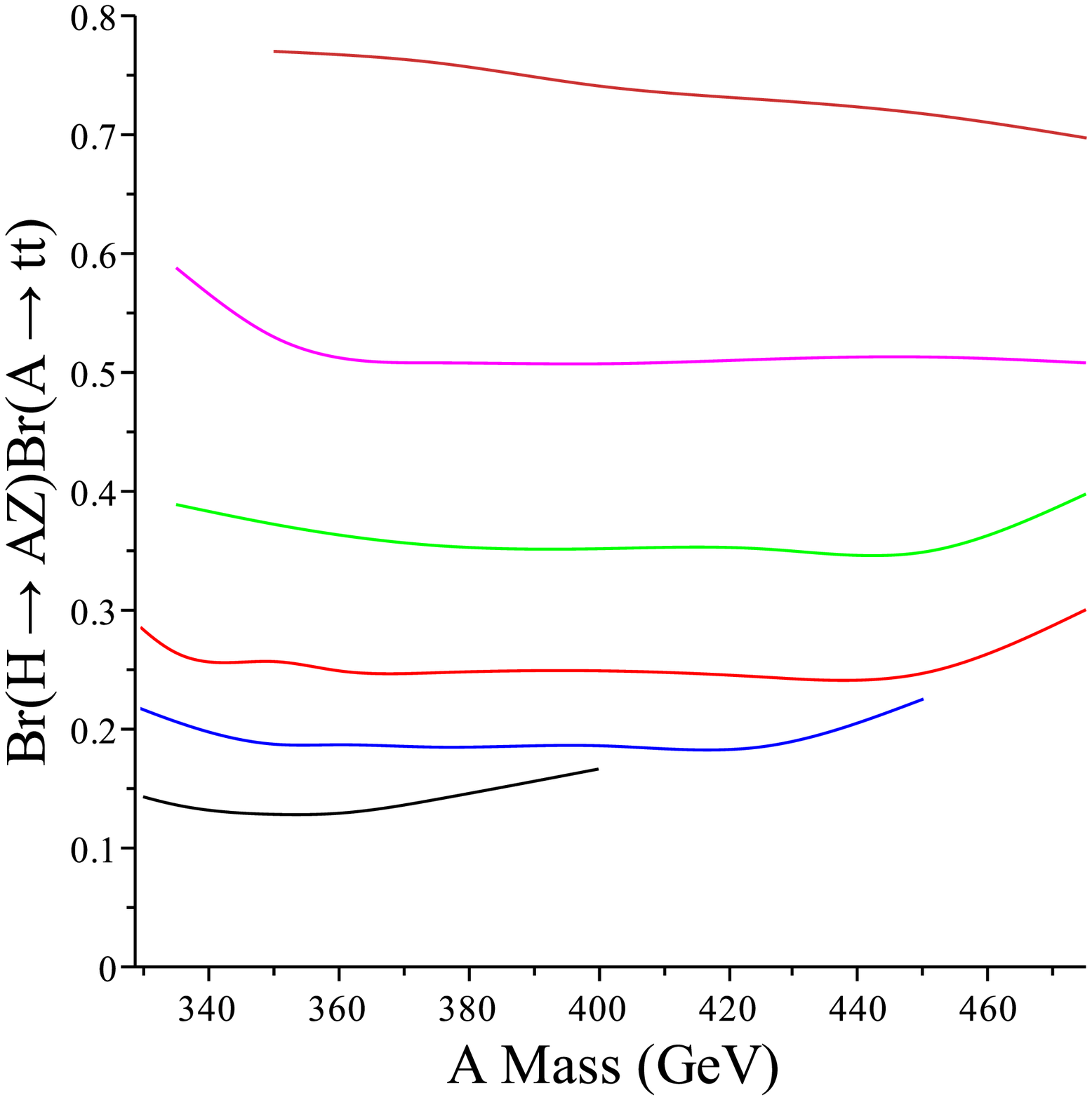}
\includegraphics[scale=0.41]{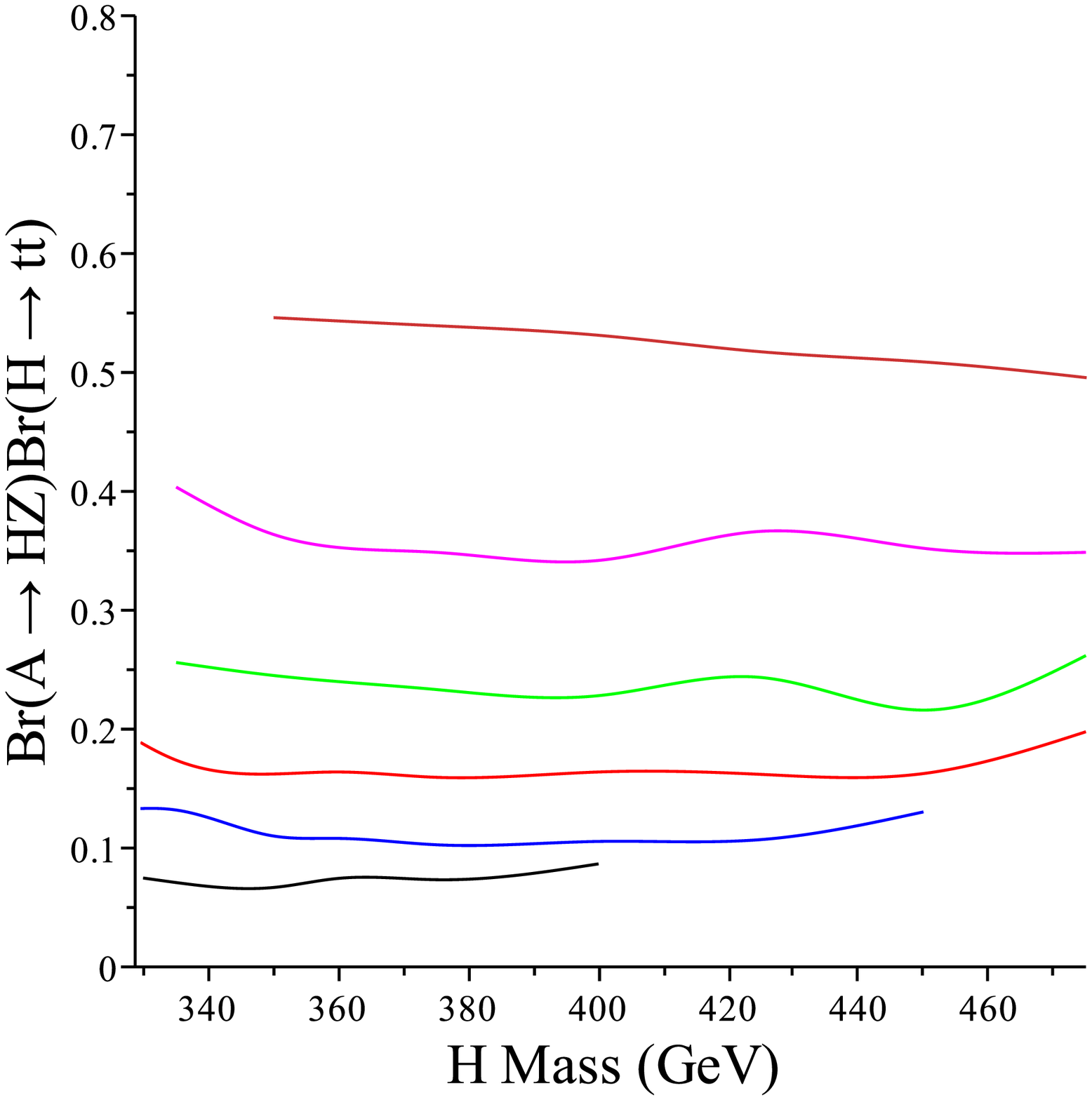}
\caption{ The upper limits on $\mathrm{Br}(H\to A Z)\mathrm{Br}(A \to t\bar{t})$ (left) and $\mathrm{Br}(A\to H Z)\mathrm{Br}(H \to t\bar{t})$ (right) as a function of $m_A$ ($m_H$) (left and right respectively). The masses of the heavier of $A/H$ are from bottom to top [500 (black),550 (blue),600 (red),650 (green),700 (magenta) and 750 (orange)] GeV. 
}
\label{ratio}
\end{figure}

For the $H$ production cross section we use the SM heavy Higgs gluon-gluon fusion 
cross section $\sigma_{SM}$  from \cite{Heinemeyer:2013tqa}.   The $A$ production cross section $\sigma_{SM}^A$ is larger than  $\sigma_{SM}$ by a mass dependent scale factor that we deduce from the results in \cite{Gunion:1989we}.  In Fig.~\ref{ratio} we show our resulting upper limits on  the product of the branching ratios $\mathrm{Br}(H/A\to A/H Z) $ and $\mathrm{Br}(A/H \to t\bar{t})$ as a function of $m_A$ and $m_H$.  These results can easily be scaled to account for different values of the production cross sections.

Our signal has similar characteristics to the SM $t\bar{t} Z$ production and if anything  our signal has a higher acceptance times efficiency. Thus we can compare our signal with the  $t\bar{t}Z$ background  for any given search and quickly estimate additional possible limits. We have not found more stringent limits from other available LHC searches.

\subsection{2HDM limits in the alignment limit}
The 2HDM is a useful benchmark model in which to discuss these decays.  In order to ensure that our results are consistent with a SM Higgs boson we work in the alignment limit where $\sin(\beta-\alpha)=1$, and we set $\lambda_6=\lambda_7=0$ \cite{Gunion:2002zf} as well. At tree level the $H/A \to A/H Z$ decay is proportional to $\sin(\beta-\alpha)$ while the decays $H\to WW,ZZ$ and $A \to hZ$ are proportional to $\cos(\beta-\alpha)$ and are thus absent in the alignment limit. The $H/A f\bar{f}$ couplings do not vanish in this limit and  $t\bar{t}$ becomes an important decay mode when  $m_{H/A}\gtrsim 2m_t$. As long as this decay dominates the other fermion decays our results are not very dependent on the type of the 2HDM. We use the Two Higgs Doublet Model Calculator \cite{Eriksson:2009ws} to calculate the branching ratios for different values of $m_{H}$ and $m_A$. We satisfy  constraints from stability of the potential, unitarity and oblique parameters, in particular by allowing $m_{12}^2$ to be  freely varying. Perturbativity is satisfied for most of the mass pairs.

\begin{figure}[htb]	
\centering
\includegraphics[scale=0.40]{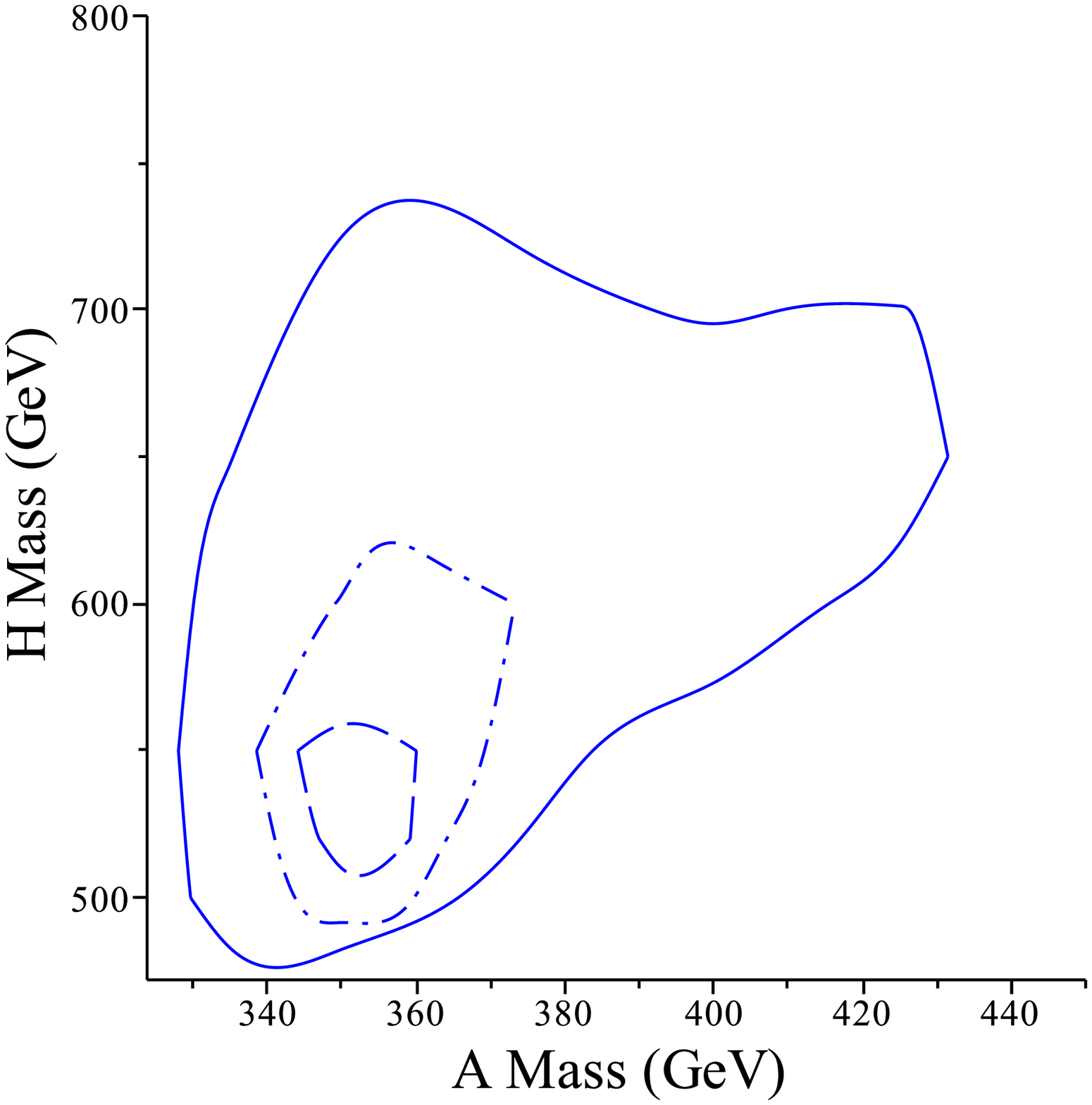}
\includegraphics[scale=0.40]{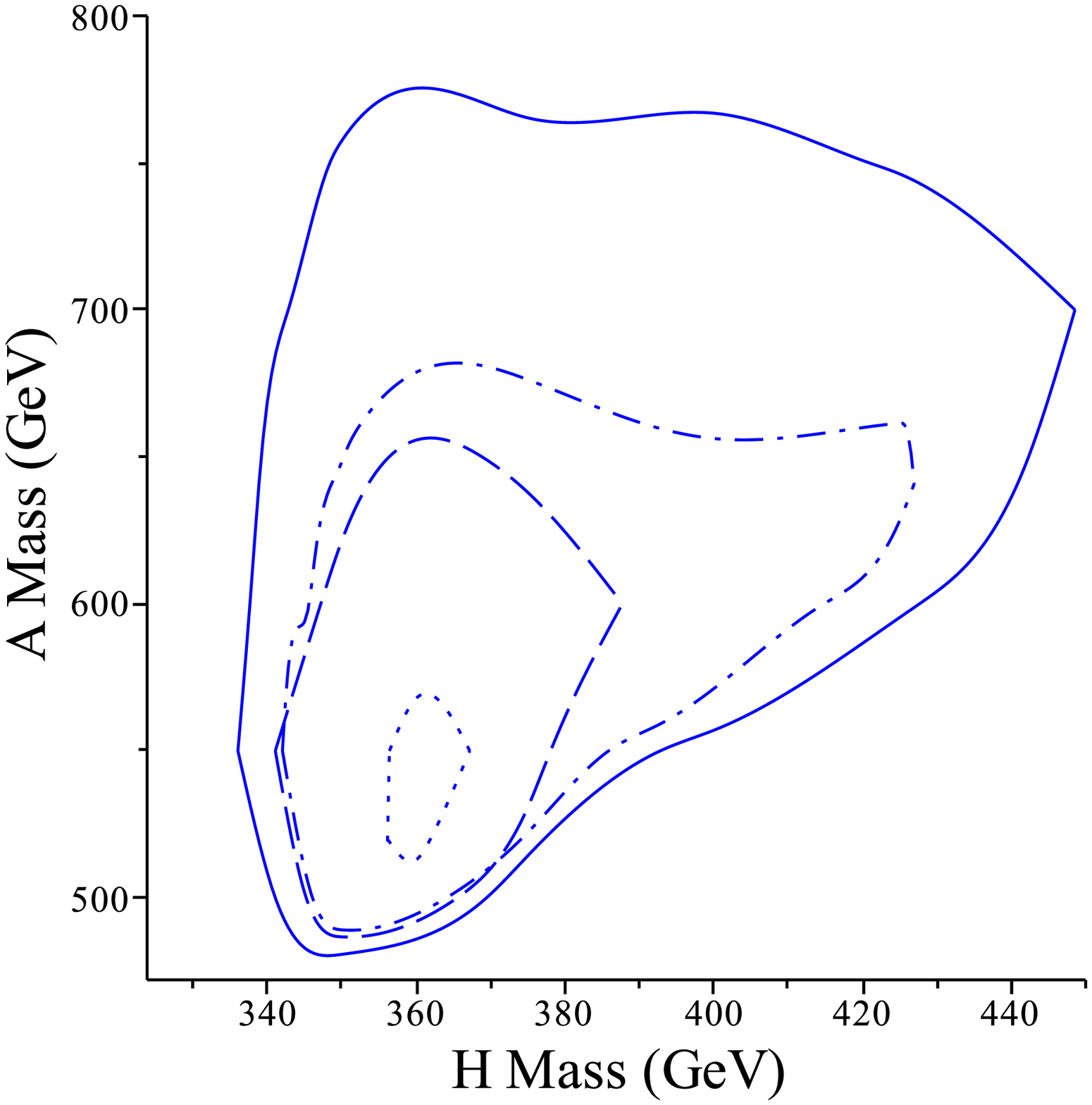}
\caption{The left (right) plot shows the mass limits for $H\to AZ$ ($A\to HZ$). The excluded region is enclosed by the solid curves for $\tan \beta =1$,  the dot-dashed curves for $\tan \beta=1.5$, and  the dotted curve for $\tan\beta= 2$ (only in the $A\to HZ$ case). The dashed curves are the limits when $m_{H^\pm}$ is the lighter of $A/H$  with $\tan \beta =1$. }
\label{ratio3}
\end{figure}

We consider both of the cases $m_H>m_{A}$ and $m_A>m_{H}$ for some low values of $\tan \beta$.  
We set  the charged Higgs mass to be the same as either the heavier or the lighter of $H$ and $A$. In the latter case there is only a limit at $\tan \beta =1$ since  the competing decay mode $H/A\to H^\pm W^\mp$ is available  with a somewhat larger  branching ratio.  
The excluded mass regions  with    $\tan \beta=\{ 1,1.5,2\}$  are shown in Fig.~\ref{ratio3}.  Constraints on the large mass are as high as $\sim750$ GeV while constraints on the mass differences are as small as $\sim130$ GeV.

For the $H^{\pm} W^{\mp}$ decay mode we consider  $A/H\to H^{\pm} W^{\mp} \to tb W\to WWbb$ which has a large background from SM $t\bar{t}$ production.   We choose  masses that are allowed from Fig.~\ref{ratio3}, $m_{H^\pm}=350$ GeV and $m_{H/A}=650$ GeV. We then scan  the relevant searches available in CheckMATE. We find that the signal contributes at most a relatively small number of events to some of the signal regions in \cite{Aad:2014mha, Aad:2014kra,Aad:2015pfx, ATLAS:2013pla}.  
  
We may use the  ATLAS  $t\bar{t}Z$ measurement at 13 TeV \cite{ATLAS13ttV} to help estimate future  13 TeV limits. Assuming the future observations are consistent with the  SM, 10 fb$^{-1}$ of data  will provide comparable limits to our $8$ TeV analysis.  The improvement from $100$ fb$^{-1}$ of data is such that the dashed and dotted curves on the corresponding plots would move out to roughly the location of the solid curves in Fig.~\ref{ratio3}. The solid curves would extend out as far as $\sim900$ and $\sim550$ GeV in the vertical and horizontal directions respectively. The constraints on the mass differences would be as small as $\sim110$ GeV.

During the LHC run 2 other types of searches will further constrain heavy Higgs masses. Heavy Higgs boson production in association with top quarks, $t\bar{t}H/A$ and $ tWb H/A $, with $ H/A\to t\bar{t}$ is another process of interest for the alignment limit at low $\tan\beta$. With $100$ fb$^{-1}$ of data at 13 TeV this process can rule out a heavy Higgs mass in the 500 to 700 GeV range at $\tan \beta =1$ (see e.g.~\cite{Craig:2016ygr}). This process does not yield a corresponding constraint from the 8 TeV data.

\section{The $750$ GeV diphoton excess from $A\to \gamma\gamma$}

Recently ATLAS \cite{ATLASgaga,ATLASgaga2} and CMS \cite{CMS:2015dxe,CMS:2016owr} have observed an excess of events in the diphoton spectrum at 750 GeV. We consider a 2HDM-like scenario  where the possible $750$ GeV state is  a pseudoscalar ($A$) that can decay to $HZ$ and $H^\pm W^\mp$. We again ignore the $H\to hh,WW,ZZ$ and $A\to hZ$ decay modes that are not present at tree level in the alignment limit. 
We assume that the 2HDM is supplemented by additional heavy states so that the 
 $A\to\gamma\gamma$ and  $gg\to A$ rates are sufficiently enhanced. Some choices that we make, such as setting  $m_H=m_{H^\pm}$, are motivated by a benchmark model that we outline below. Within this set of assumptions we consider constraints coming from our  $t\bar{t}Z$  analysis  and the CMS $t\bar{t}$ resonance limits \cite{Chatrchyan:2013lca}. We consider   $m_H\geq 550$ GeV and the $A\to t\bar{t}$ partial width  in the range $\Gamma^A_{t\bar{t}}=0$\textendash40 GeV. The upper end of this range is the natural width in the 2HDM at $\tan \beta= 1$.

The partial widths to $HZ$ and $H^\pm W^\mp$ at the masses that  we consider are shown in Table \ref{BR}, as are our limits on the cross section for the production and decay of $A$ ending with $t\bar{t}Z$. 
\begin{table}[htp]
\begin{center} 
\begin{tabular}{|c|ccccc|}
\hline
$m_{H}=m_{H^\pm}$ GeV&$550$&$600$ &$625$ &$640$ &$650$ \\
\hline
 $\Gamma^{A}_{HZ} $ GeV&$9.6$&$3.2$&$1.3$&$0.48$&$0.15$\\
   $\Gamma^{A}_{H^\pm W^{\mp}}$ GeV&$21$&$7.7$&$3.5$&$1.8$&$0.9$\\
   Limit $\sigma_{t\bar{t} Z}$ fb&$106$&$114$&$117$&$132.5$&$147$\\
\hline
\end{tabular}
\end{center}
\caption{The partial widths for $A\to HZ$ ($\Gamma^{A}_{HZ}$)  and $A\to H^\pm W^\mp$ ($\Gamma^{A}_{H^\pm W^{\mp}}$). }
\label{BR} 
\end{table}

We supplement the limits from our $t\bar{t}Z$ analysis by limits on the $A\to t\bar{t}$ decay. The CMS $8$ TeV search for $t\bar{t}$
 resonances  \cite{Chatrchyan:2013lca} sets un upper limit of $\sigma_{t\bar{t}}<0.3$ pb on a narrow scalar  resonance at $750$ GeV. 
 The corresponding ATLAS limit is $\sigma_{t\bar{t}}<0.7$ pb \cite{Aad:2015fna}.   Neither ALTAS nor CMS has a search for a wide width scalar $t\bar{t}$ resonance. CMS \cite{Chatrchyan:2013lca} has a constraint on wide $t\bar{t}$ resonance due to a $750$ GeV $Z'$, where $\sigma_{Z'}<512$ fb for a $10\%$ width. This provides a rough estimate for the wide width scalar limit. 

The $\gamma\gamma$ cross section should be  $\sim 4$\textendash9 fb \cite{Buckley:2016mbr} in order to account for observed excess at $13$ TeV. The corresponding $8$ TeV cross section is $\sim 0.6$\textendash2 fb. There is also  an upper limit of $\sim 1.3$ fb from the CMS  $8$ TeV data \cite{Khachatryan:2015qba} while the ATLAS $8$ TeV limit is weaker \cite{Aad:2015mna}.  We use $\sigma_{\gamma\gamma}\sim 0.9$ fb at $8$ TeV to accommodate the CMS limit and the $13$ TeV excess.  This value corresponds to the narrow width best fit value of $\sim 4$ fb at $13$ TeV \cite{Buckley:2016mbr}.

We express these constraints on the $\gamma\gamma$, $t\bar{t}$ and $t\bar{t}Z$ cross sections in terms of the partial widths for $A\to gg $ ($\Gamma_{gg}^A $) and $A\to \gamma\gamma $ ($\Gamma_{\gamma\gamma}^A $). We have
 \begin{align}
\sigma_{\gamma\gamma}&=\frac{\Gamma^A_{\gamma\gamma}}{\Gamma_{tot}}\sigma^A_{prod} =\frac{\Gamma^A_{\gamma\gamma}}{\Gamma_{tot}}\frac{\Gamma^A_{gg}}{\Gamma^{A _{SM}}_{gg}}\sigma^A_{SM},\\
\Gamma_{tot}&=\Gamma^A_{t\bar{t}}+\Gamma^A_{HZ}+\Gamma^A_{H^\pm W^\mp}+\Gamma_{gg}^A+\Gamma^A_{\gamma\gamma}+\Gamma_{other}.
\label{gaga cross sections}
\end{align}
 The largest contributions to $\Gamma_{other}$ are from $b\bar{b}$ and $\tau\bar{\tau}$.
The production cross section $\sigma^A_{prod}$ is enhanced due to the loop contributions of new particles as reflected in the value of $\Gamma^A _{gg}$.\footnote{The remaining ratio is $\sigma^A_{SM}/\Gamma^{A _{SM}}_{gg}\sim 5\times 10^3 \mbox{ fb/GeV}$ where we have used the same cross section  $\sigma^A_{SM}$ as in the first section.}
The observed value of $\sigma_{\gamma\gamma}$ determines contours of fixed $\Gamma^{A}_{t\bar{t}}$ as a function of $ \Gamma^A_{gg}$ and $\Gamma_{\gamma\gamma} ^A $. The basic effect of the $HZ$ and $H^\pm W^\mp$ decay modes is to push up the required value of the product $\Gamma^A_{\gamma\gamma}\Gamma^A_{gg}$.

\begin{figure}[htb]	
\centering
\includegraphics[scale=0.29]{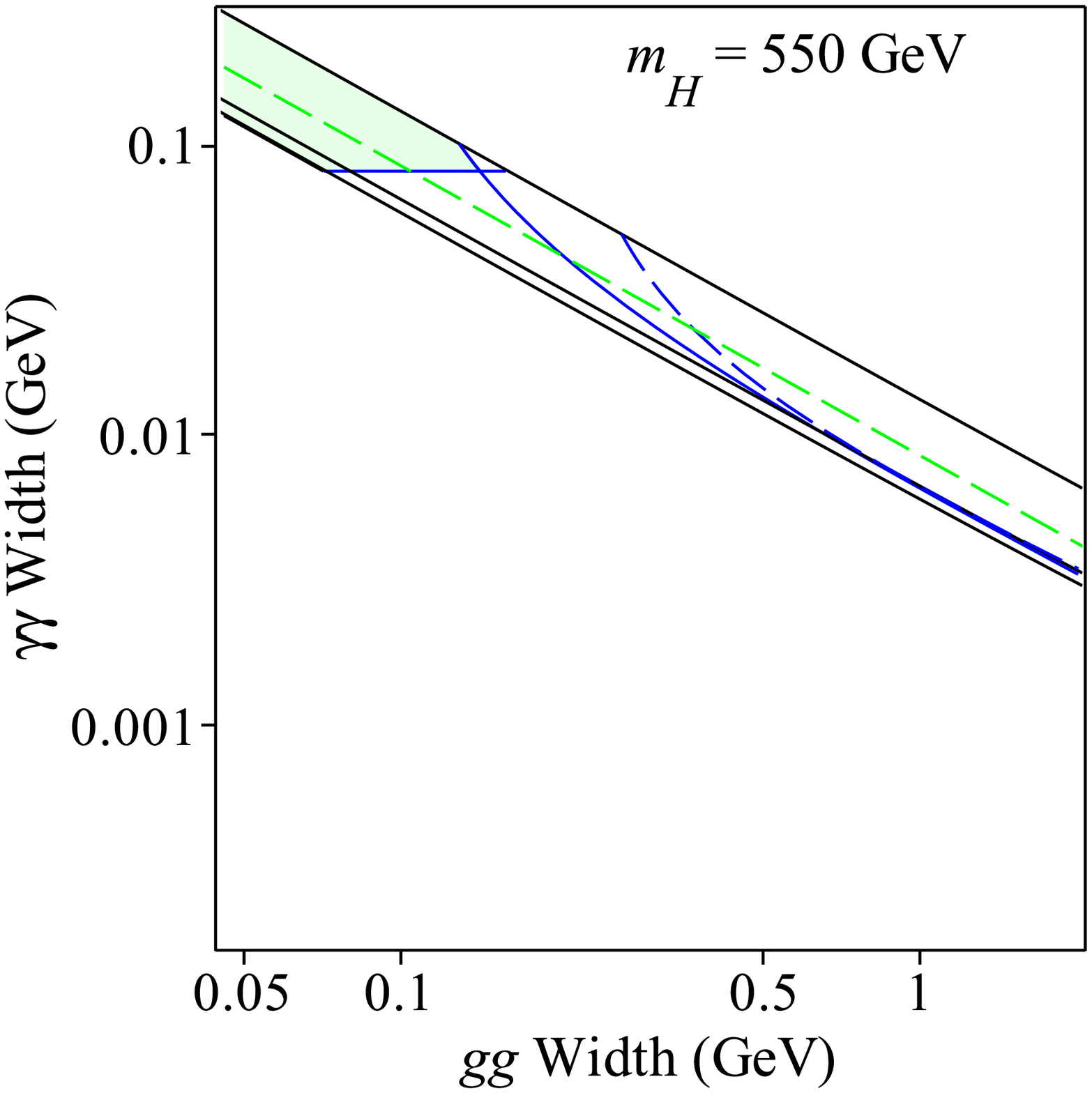}
\includegraphics[scale=0.29]{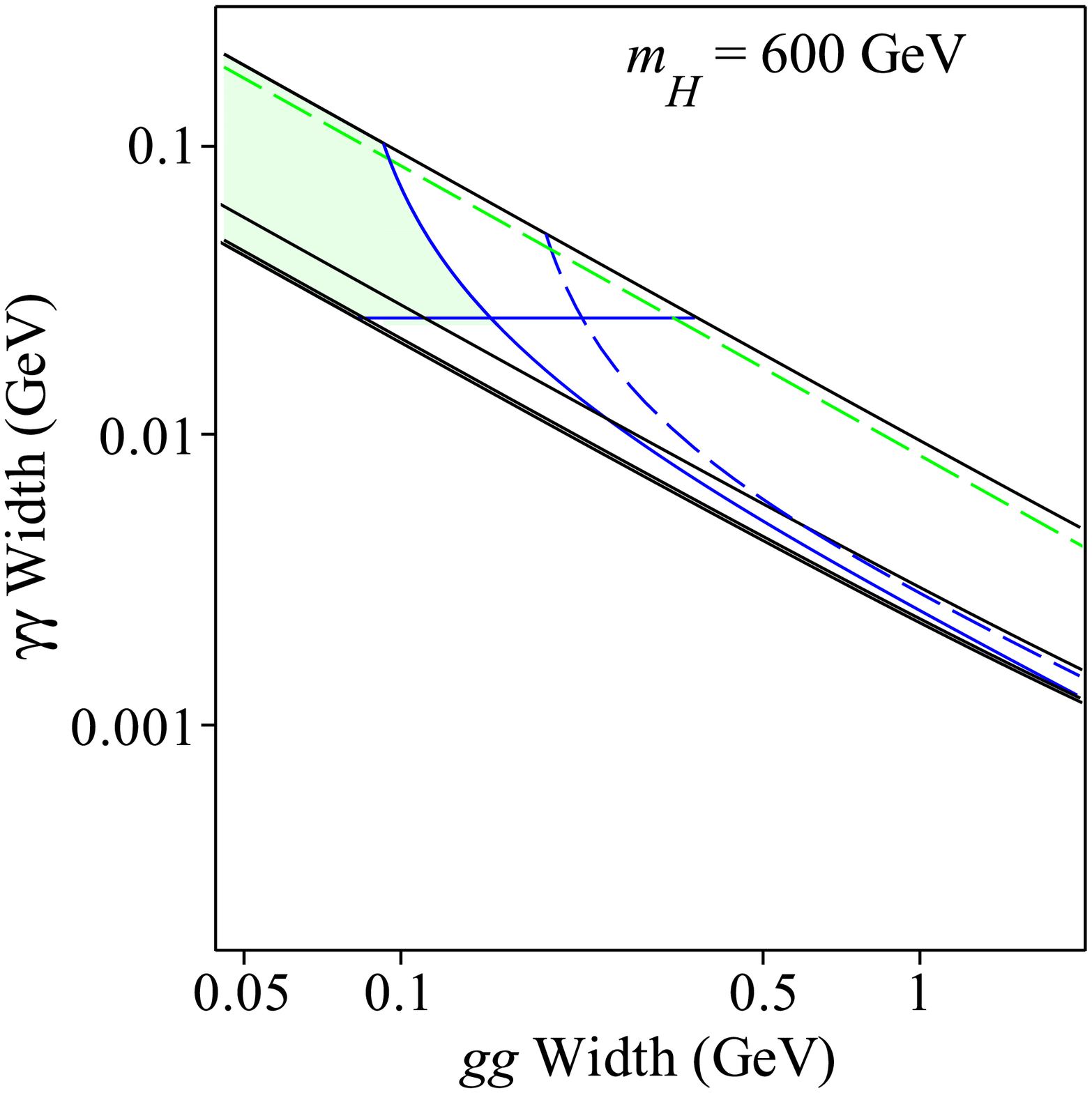}
\includegraphics[scale=0.29]{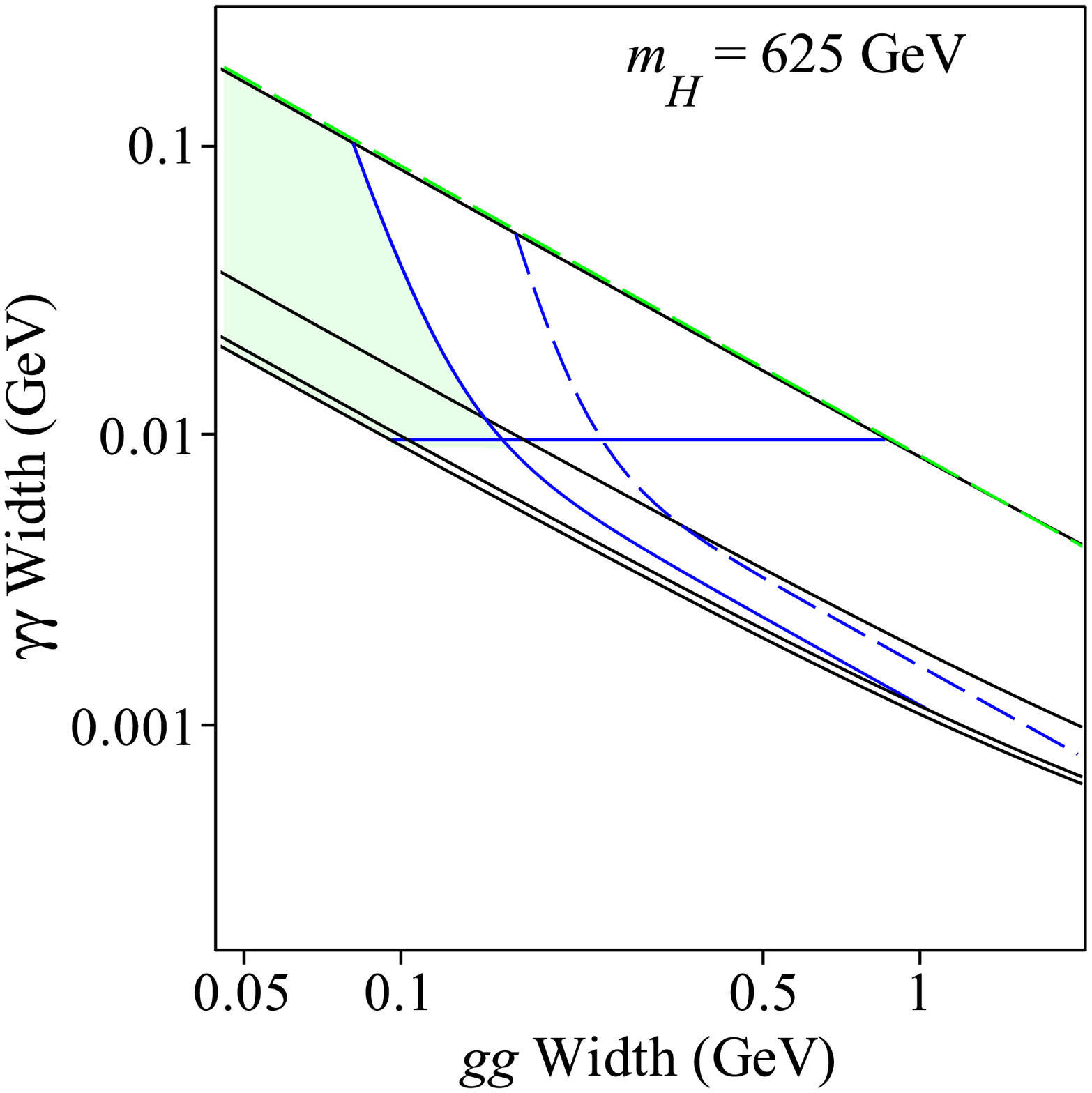}
\includegraphics[scale=0.29]{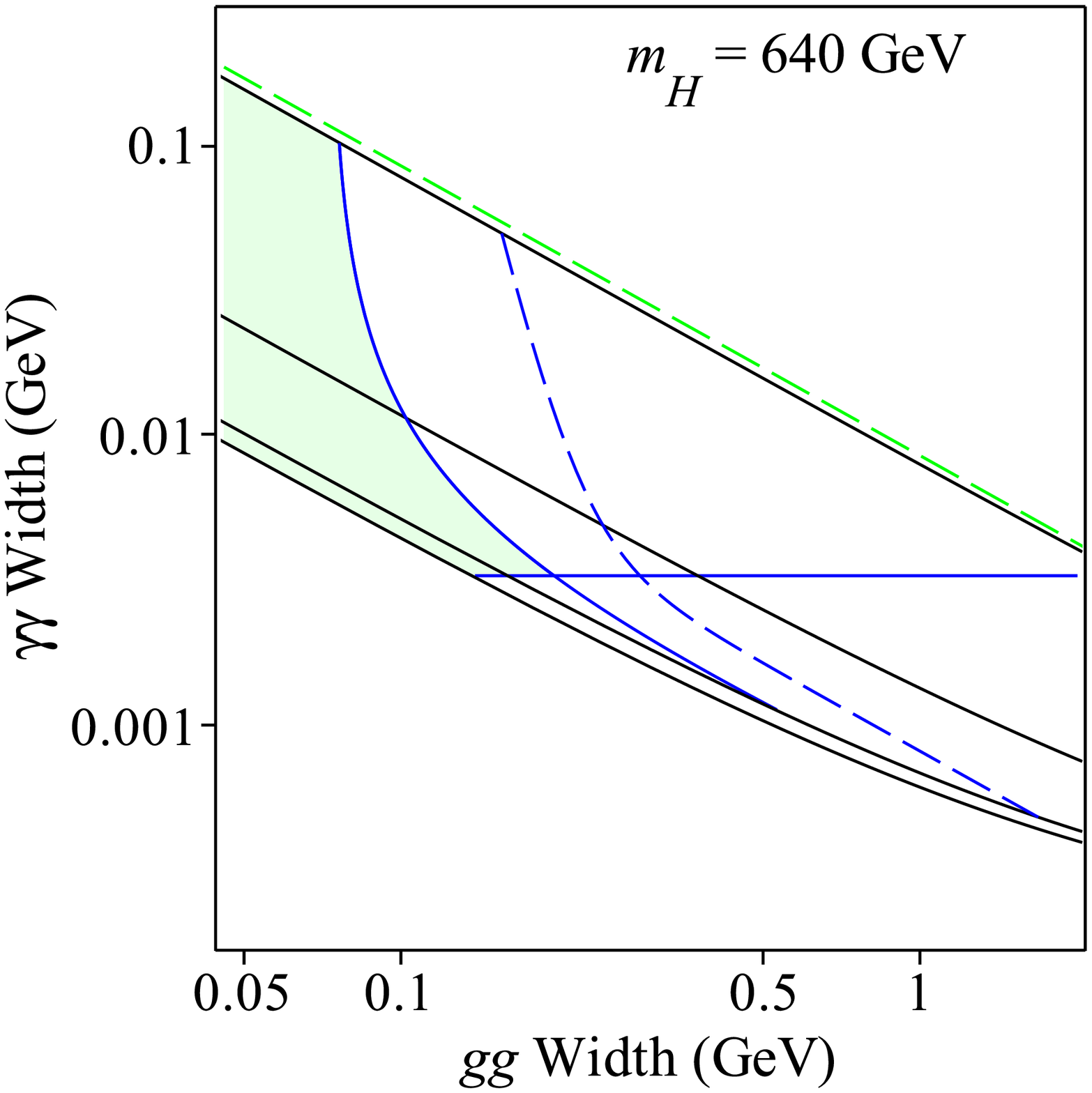}
\includegraphics[scale=0.29]{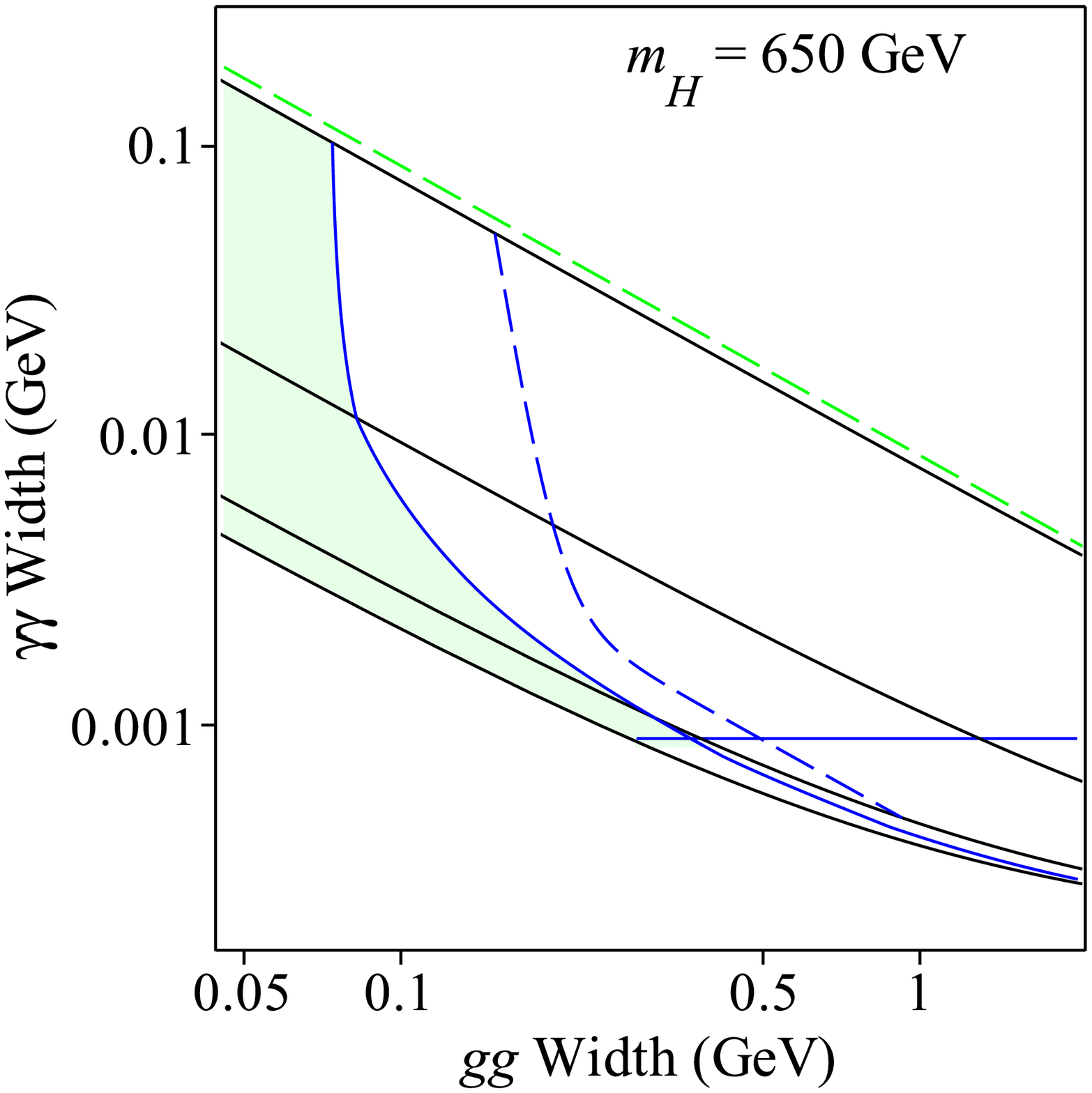}
\includegraphics[scale=0.29]{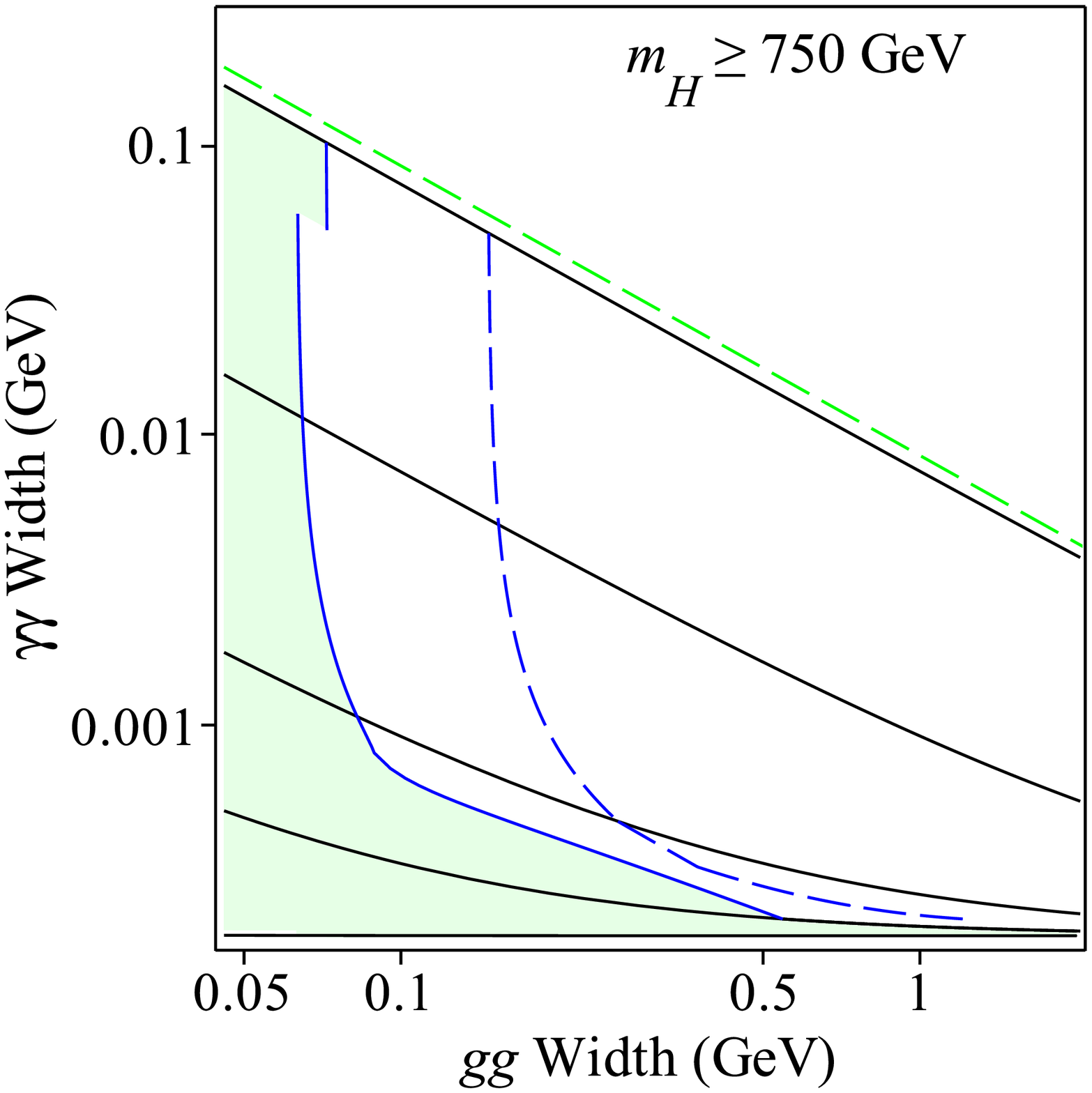}
\caption{The green shaded region shows the allowed values of the $\gamma\gamma$ and $gg$ widths consistent  with $\sigma_{ \gamma \gamma}\sim 0.9$ fb at 
$8$ TeV. The blue solid lines are the limits from  $t\bar{t} $  (curved line)  and  $t\bar{t}Z$ (horizontal line). The dashed blue line is the ATLAS $t\bar{t} $ limit. Fixed $t\bar{t}$ widths correspond to the black curves where
from bottom to top  $\Gamma^{A}_{t\bar{t}}\sim 0,0.4,4,40$ GeV. The dashed green lines indicate a total width of $\sim 45$ GeV. The $m_H \gtrsim 750$ GeV case lacks a smooth continuation from the wide width to the narrow width limits, and for this case we have inserted an additional $\Gamma^{A}_{t\bar{t}}\sim 0.08$ GeV curve.  }
\label{masterplot1}
\end{figure}

 The limits on the cross sections $\sigma_{t\bar{t}}$ and $\sigma_{t\bar{t}Z}$ can be written in terms of the partial widths as follows:
  \begin{align}
\sigma_{t\bar{t}}
&=\sigma_{\gamma\gamma}\frac{\Gamma^A_{t\bar{t}}}{\Gamma^A_{\gamma\gamma}},\\
\sigma_{t\bar{t}Z}
&=\sigma_{\gamma\gamma}\frac{\Gamma^A_{HZ}\mathrm{Br}(H\to t\bar{t})}{\Gamma^A_{\gamma\gamma}}.
\label{limit cross sections 2}
\end{align}
The upper limits on $\sigma_{t\bar{t}}$ and $\sigma_{t\bar{t}Z}$ then determine the allowed regions on our plots. As can be seen from \eqref{limit cross sections 2} and the results in Table \ref{BR} the $t\bar{t}Z$ limit simply  sets a $m_H$ dependent minimum value
for $\Gamma_{\gamma\gamma}^A$. [We expect that $\mathrm{Br}(H\to t\bar{t})$ is close to unity.] The allowed regions and the $\Gamma^A_{t\bar{t}}$ contour curves  are shown  in Fig.~\ref{masterplot1}. 

 The maximum allowed $gg$  and the minimum allowed $\gamma\gamma$ widths occur at the intersection point of the two limit curves in Fig.~\ref{masterplot1} and they are shown in Table \ref{limit points}. Different possible choices of $\sigma_{\gamma\gamma}$ at $13$ TeV only affect the minimum $\gamma\gamma$ widths as indicated in the table.  
 \begin{table}[htp]
\begin{center} 
\begin{tabular}{|c|ccccc|}
\hline
$m_{H}=m_{H^\pm}$ GeV&$550$&$600$ &$625$ &$640$ &$650$ \\
\hline
  Max $\Gamma^{A}_{gg}$ GeV&$0.14$&$0.15$&$0.16$&$0.185$&$0.36$\\
   $\Gamma_{t\bar{t}}^A$ GeV&$32$&$8.3$&$3.3$&$1.3$&$0.31$\\
   \hline
 Min $\Gamma^{A}_{\gamma\gamma} $ GeV  at  $\sigma_{\gamma\gamma}\sim6$ fb&$0.12$&$0.037$&$0.014$&$0.005$&$0.0013$\\
 Min $\Gamma^{A}_{\gamma\gamma} $ GeV  at     $\sigma_{\gamma\gamma}\sim5$ fb&$0.1$&$0.031$&$0.012$&$0.004$&$0.0011$\\
 Min $\Gamma^{A}_{\gamma\gamma} $ GeV  at  $\sigma_{\gamma\gamma}\sim4$ fb&$0.08$&$0.025$&$0.01$&$0.003$&$0.0009$\\
   Min $\Gamma^{A}_{\gamma\gamma} $ GeV  at  $\sigma_{\gamma\gamma}\sim3$ fb &$0.06$&$0.019$&$0.007$&$0.0024$&$0.00067$\\
\hline
\end{tabular}
\end{center}
\caption{The minimum (maximum) of the $\gamma\gamma$ ($gg$) partial width for various choices of $\sigma_{\gamma\gamma}$ at $13$ TeV. The corresponding value for the $t\bar t$ partial width is also shown.}
\label{limit points} 
\end{table}
 
\subsection{Benchmark model}

 In order to provide some background for the choices we have made we briefly describe   a  benchmark model. Here the new heavy states are a chiral quark doublet $\{t',b'\}$ and a chiral lepton doublet $\{\tau',\nu'\}$ with SM charges. How such a extension of the SM is consistent with properties of the light Higgs boson is discussed in \cite{Holdom:2014bla}. A four Higgs doublet model is thought to emerge as an effective description of the fluctuations around the chiral condensates of the four flavors $t,t',b',\tau'$.  The lighter of these bosons have a flavor content dominated by $t'$ and $b'$ and they form a set that is 2HDM-like. For these states the $H,H^\pm$ form an isotriplet and the $A$ is an isosinglet relative to an approximate custodial symmetry in the $t',b'$ sector. This implies that $m_H\approx m_{H^\pm}$. It also implies that while $A$ has an enhanced production from gluon fusion, $H$ has a suppressed production. We set $m_{t'}= m_{b'}= 800$ GeV and choose $m_{\tau'}>m_A/2$ to avoid the $A\to\tau'\tau'$ decay. For the $A$ widths of most interest to the $\gamma\gamma$ signal we find $\Gamma_{gg}^A\sim 0.25$ GeV and $\Gamma_{\gamma\gamma}^A\sim 0.0015$\textendash0.004 GeV, corresponding to $m_{\tau'}=700$\textendash375 GeV.
 
From our results in Fig.~\ref{masterplot1} we see that $m_H$ around 650 GeV can be consistent with these widths, namely the lower right corner of the allowed region in the $m_H=650$ GeV plot. But then in addition we see that the top width $\Gamma^A_{t\bar{t}}$ needs to be reduced from its naive 2HDM value of $\sim40$ GeV. The $At\bar{t}$ coupling, along with other couplings of the bosons and their mass spectrum, is ultimately determined by the multi-Higgs potential. We find that parameters of this potential, partially already constrained by the light Higgs properties, can be constrained further such that $\Gamma^A_{t\bar{t}}$ falls in the allowed range, e.g. $\Gamma^A_{t\bar{t}}\lesssim 1$ GeV.\footnote{The $A\tau'\tau'$ coupling need not be so suppressed and it tends to enhance $\Gamma_{\gamma\gamma}^A$. The $Ht\bar t$ coupling is suppressed, but still the $t\bar t$ decay of $H$ dominates.} From the plot we also see how the $t\bar t$ resonance searches of CMS and ATLAS differ on determining what this allowed range is.  For larger $m_H$, the $m_H \gtrsim 750$ GeV plot shows that the model parameters would have to be more tuned to obtain a sufficiently small $\Gamma^A_{t\bar{t}}$.
 
 \subsection{$H^\pm W^\mp$ decays and leptonic decays}
We turn now to the question of whether there can be direct limits due to the decay $A\to H^\pm W^\mp$. We consider the $A\to Wtb$ decay and find an approximate
upper limit  from \cite{Aad:2014kra} using CheckMATE. At $m_{H^\pm}=650$ GeV we estimate that  the cross section limit is $\sim 1$ pb for this final state, to be compared with the predicted value of $\sigma^A_{H^\pm W^\mp}\sim 600$ fb when $Br(H^\pm\to tb)=1$.\footnote{When $m_{H^\pm}=m_{H}=650$ GeV, $\Gamma^A_{H^\pm W^\mp}\approx 6\Gamma^A_{HZ}$.} We note that if the width of the $H^\pm$ is large (for example $\sim 50$ GeV),  the branching ratio to $A\to H^\pm W^{\mp}$ where the $H^\pm$ is virtual can remain quite unsuppressed  beyond $m_H>660$ GeV. When this is the case then for such $H$ masses the corresponding plot of the type in Fig.~\ref{masterplot1} can continue to look like the $m_H=650$ GeV plot (but without the limit from $t\bar tZ$).

In the benchmark model another possible decay mode is $H^\pm\to\tau'\nu'$, and when kinematically open it can be a dominant decay mode that results in multilepton final states. We take the dominant heavy lepton decays to be $\tau'\to W\nu'$ (rather than $\tau'\to W\nu_{\tau}$) and $\nu'\to \tau W$ ($\nu'\to \mu W, eW$ would be even more striking).  Then there is a quite striking final state,
\begin{equation}
  A\to H^\pm W^\mp\to W\tau'\nu'\to WWWW\tau \tau.  
\end{equation}
The cross section limits for this process from the  ATLAS  four  or more lepton search  \cite{ATLAS:2013qla} are approximately $ \lesssim 55$\textendash90 fb when $m_{\tau'}=400$ GeV and $m_{\nu'}=100$\textendash200 GeV. Given the size of $\sigma^A_{H^\pm W^\mp}$ and any sizable $H^\pm\to\tau'\nu'$ branching ratio, these data imply that the $\nu'$ and $\tau'$ masses are such that this cascade decay is either kinematically forbidden or at least kinematically suppressed.

\section{Summary}

When the masses of additional Higgs bosons are not degenerate, the $A/H\to t\bar{t}Z$ final state offers a relatively model independent and clean probe of the heavy Higgs sector. In the 2HDM (at low $\tan\beta$ and close to the alignment limit) we found that it eliminates quite a large region of the $m_H$\textendash$m_A$ plane. In general any future search that is sensitive to  the $t\bar{t}Z$ standard model background  may also  be sensitive to this signal. 
 
Assuming that a 750 GeV mass $A$ is the source of the diphoton excess observed by ATLAS and CMS \cite{ATLASgaga,ATLASgaga2,CMS:2015dxe,CMS:2016owr}, we considered the impact of a lighter $H$ that leads to the decay $A\to HZ\to t\bar{t}Z$.  Basically we find that the lighter the $H$, the larger the $\gamma\gamma$ width of the $A$ needs to be. In a benchmark model where the new fermions are $t',b',\tau',\nu'$ and the $\gamma\gamma$ width cannot be arbitrarily large, we find that $m_H$ is constrained  to a narrow range around $650$ GeV. In this case the model predictions nearly saturate the bounds both from the $t\bar t Z$ SM search and the $t\bar t$ resonance search. The $Wtb$ signal from the charged Higgs decay may also be of interest. Thus this picture should be quite testable as new LHC data emerge.

\begin{acknowledgments}
This research is supported in part by the Natural Sciences and Engineering Research
Council of Canada.
\end{acknowledgments}

\end{document}